\documentclass[english,aps,pra,show pacs,two column,groupedaddress]{revtex4}
\usepackage[T1]{fontenc}
\usepackage[latin9]{inputenc}
\usepackage{graphicx,array,tabularx}
\usepackage{amstext}
\usepackage{hyperref}
\usepackage{bm,amssymb,amsmath,verbatim,cleveref}



\usepackage{babel}
\usepackage{color}

\begin{document}
\title{Characterization of decohering quantum systems: Machine learning approach}
\author{Markku P.V. Stenberg}
\email[]{markku.stenberg@iki.fi}
\author{Oliver K{\"o}hn}
\email[]{oliver@alef.lusi.uni-sb.de}
\author{Frank K. Wilhelm}
\affiliation{Theoretical Physics, Saarland University, 66123 Saarbr{\"u}cken, Germany}


\begin{abstract}
Adaptive data collection and analysis, where data are being fed back to update the measurement settings, can greatly increase speed, precision, and reliability of the characterization of quantum systems. However, decoherence tends to make adaptive characterization difficult. As an example, we consider two coupled discrete quantum systems. When one of the systems can be controlled and measured, the standard method to characterize another, with an unknown frequency $\omega_{\rm r}$, is swap spectroscopy. 
Here, adapting measurements can provide estimates whose error decreases exponentially in the number of measurement shots rather than as a power law in conventional swap spectroscopy. However, when the decoherence time is so short that an excitation oscillating between the two systems can only undergo less than a few tens of vacuum Rabi oscillations, this approach can be marred by a severe limit on accuracy unless carefully designed. We adopt machine learning techniques to search for efficient policies for the characterization of decohering quantum systems. We find, for instance, that when the system undergoes more than 2 Rabi oscillations during its relaxation time $T_1$, $O(10^3)$ measurement shots are sufficient to reduce the squared error of the Bayesian initial prior of the unknown frequency $\omega_{\rm r}$ by a factor $O(10^4)$ or larger. We also develop policies optimized for extreme initial parameter uncertainty and for the presence of imperfections in the readout.
\end{abstract}
\maketitle
\section{Introduction}
Maintaining coherence poses a major challenge for quantum information processing. During the last fifteen years, coherence times of quantum bits have increased dramatically, for instance, for superconducting qubits by a factor $\sim 10^{5}$, from a few ns \cite{nakamura99} to hundreds of $\mu$s \cite{devoret13}. Nevertheless, decoherence remains the main source of errors in quantum circuits and enhances the resource requirements for fault-tolerant quantum computation \cite{fowler12}. In addition to the control of the quantum circuits, decoherence also makes characterizing them more difficult. Accurate characterization of quantum devices is becoming increasingly important as the requirements for their precise operation become more stringent.

The most complete, and in larger systems, inhibitively resource intensive, approach to characterization is full process tomography \cite{chuang97}. A more feasible technique is to obtain a single quantity describing the level of coherent control, {\it e.g.}, fidelity, through randomized benchmarking \cite{emerson05,knill08,magesan11,magesan12}. In addition to these, a 
third methodology, based on adaptive measurements, has recently advanced strongly both theoretically \cite{berry09,sergeevich11,huszar12,granade12,ferrie13,wiebe14b,wiebe14a,stenberg14,stenberg15} and experimentally \cite{higgins07,xiang11,yonezawa12,kravtsov13}.
Rather than choosing measurement settings prior the experiment, the latter approach takes advantage of the data already while it is collected and chooses measurements based on the
current knowledge and uncertainty of the parameters. The main task in characterizing a particular system is to assign a set of rules, also called a policy, according to which the measurement settings can be chosen efficiently based on the data obtained.
 
Adapting measurements is a generally powerful approach that may deliver performance that is not achievable through nonadaptive schemes, as becomes particularly apparent in inferring several unknown system parameters \cite{wiebe14b,wiebe14a,stenberg14,stenberg15}. Often the advantages of adaptive measurement schemes are also robust against moderate experimental imperfections \cite{wiebe14b,wiebe14a,stenberg14,stenberg15}. We note, however, that loss of coherence can pose significant limitations on the achievable speed-up. Indeed, as we will demonstrate below, in practice, short relaxation time often leads to loss of the numerical accuracy in adaptive estimation unless the corresponding policies are carefully designed.

In this paper we, rather than merely demonstrating robustness against certain error rates, present policies tailored for different degrees of energy relaxation, readout error rate and initial parameter uncertainty. While searching such policies through manual work would be inhibitively time consuming, we show it can be conveniently done through a machine learning algorithm that is able to improve policies based on past experience. For definiteness, we characterize an initially unknown frequency $\omega_{\rm r}$ of a mode coupled to a qubit with a controllable frequency $\omega_{\rm q}$ through a coupling with an unknown strength $g$. However, similar ideas are applicable more generally when waiting time $t$ between the preparation and the measurement of the system is one of the relevant control parameters.
\section{Experimental resources}
A qubit coupled to another mode is described by the Jaynes-Cummings Hamiltonian \cite{blais04}
\begin{equation}
\hat{H}_{{\rm JC}}=\frac{\hbar\omega_{\rm q}}{2}\hat{\sigma}_{z}+\hbar\omega_{\rm r}\left(\hat{a}^{\dagger}\hat{a}+\frac{1}{2}\right)+\hbar g\left(\hat{\sigma}_{+}\hat{a}+\hat{\sigma}_{-}\hat{a}^{\dagger}\right)
\end{equation}
where $g\ll\omega_{\rm r}$. Here $\omega_{\rm q}$ is the qubit frequency and $\omega_{\rm r}$ the frequency of another mode, {\it e.g.}, in a stripline resonator or a spurious 
two-level system (TLS) in the junction. 

The conventional method to estimate $g$ and $\omega_{\rm r}$ is called swap spectroscopy \cite{mariantoni11,swap}. One first prepares the qubit in its excited state and the cavity in the ground state. The frequency $\omega_{\rm q}$ is then fixed to a chosen value after which the system is allowed to evolve a time $t$ during which the excitation undergoes vacuum Rabi oscillations between the qubit and the resonator. The qubit is then measured in the $\sigma_z$ basis. The system is thereafter reset to its ground state before the next measurement. When the qubit may undergo energy relaxation, its  ground state occupation probability oscillates around an increasing envelope
\cite{stenberg14} 
\begin{align}
&P_{\omega_{\rm q},t}\left(0|g,\omega_{\rm r}\right)=1-\left(\frac{\omega_{\rm R}+\Delta\omega}{2\omega_{\rm R}}\right)^{2}\exp\left[\frac{-(\omega_{\rm R}+\Delta\omega)t}{2\omega_{\rm R}T_1}\right] \nonumber \\&-\left(\frac{\omega_{\rm R}-\Delta\omega}{2\omega_{\rm R}}\right)^{2}\exp\left[\frac{-(\omega_{\rm R}-\Delta\omega)t}{2\omega_{\rm R}T_1}\right]\nonumber \\&-\frac{2g^{2}}{\omega_{\rm R}^{2}}\exp\left(\frac{-t}{2T_1}\right)\cos \omega_{\rm R}t,
\end{align}
where $\Delta\omega=\omega_{\rm q}-\omega_{\rm r}$ is the detuning frequency, $T_1$ the relaxation time of the qubit, and $\omega_{\rm R}=\sqrt{\Delta\omega^2+4g^2}$. 
In the absence of relaxation this simplifies to
\begin{eqnarray}
&P_{\omega_{\rm q},t}\left(0|g,\omega_{\rm r}\right)=\frac{1}{2}\left(1-\frac{4g^{2}}{\omega_{\rm R}^{2}}\cos\omega_{\rm R}t-\frac{\Delta\omega^{2}}{\omega_{\rm R}^{2}}\right).
\end{eqnarray}
In conventional swap spectroscopy, the measurement is repeated at a setting $(\omega_{\rm q},t)$ many, usually thousands of times, in order to accurately approximate the probability 
of the qubit to be in its ground state. The patterns in this probability surface as a function of $(\omega_{\rm q},t)$ are then used to obtain the estimate $(\tilde{g},\tilde{\omega}_{\rm r})$.

Most rapidly improving estimates can be found by preparing the initial state as a superposition state and by performing the qubit measurements in an adaptive basis (rather than in a fixed $\sigma_z$ basis) \cite{wiebe14a,wiebe14b,yuan15}. But here we assume more modest experimental resources, with $(\omega_q,t)$ as the only experimental control knobs, {\it i.e.}, the setup of standard swap spectroscopy only enhanced by the ability to update $(\omega_q,t)$ adaptively. We show that even this setup allows rather efficient characterization.
\section{Bayesian inference}
\begin{figure}[h!]
\includegraphics[width=0.5\textwidth]{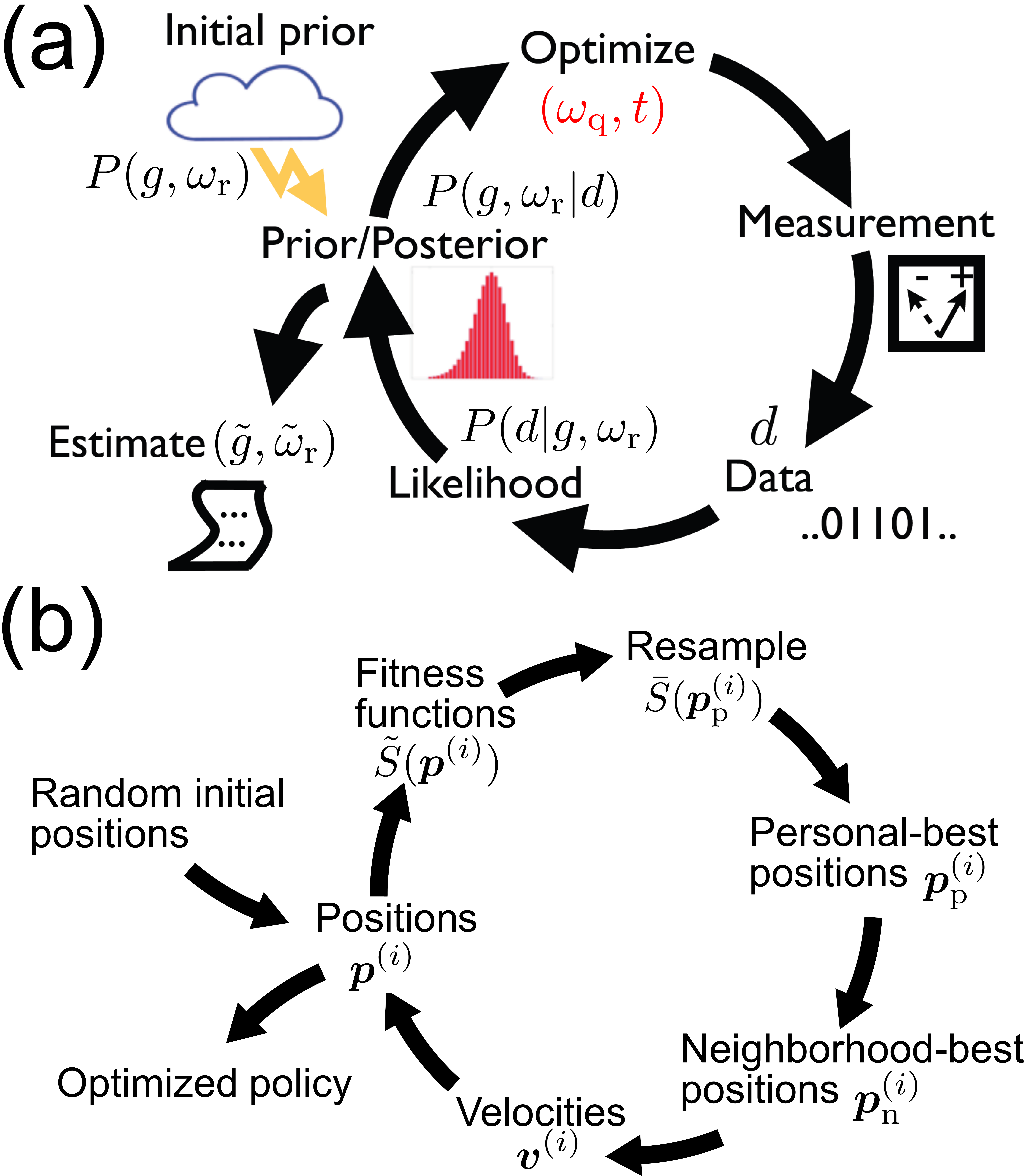}
\caption{(Color online) Schematics of (a) an adaptive Bayesian inference scheme and (b) particle swarm optimization algorithm.}
\label{fig:fig1}
\end{figure}

Our policy does not attempt to accurately estimate the ground state occupation probability of the qubit through averaging over large ensembles. Instead, we extract the estimates of the unknown parameters by making use of Bayes' theorem
\begin{equation}
P(g,\omega_{\rm r}|d)=\frac{P(d|g,\omega_{\rm r})P(g,\omega_{\rm r})}{P(d)}.
\label{eq:bayes_theorem}
\end{equation}
The starting point is the initial prior probability distribution $P(g,\omega_{\rm r})$ that quantifies one's {\it a priori} conception about the unknown parameters and the degree of their uncertainties. Having obtained data $d$, one calculates the likelihood function $P(d|g,\omega_{\rm r})$, {\it i.e.}, the probability to obtain that data with different parameter values prior the measurement.  The normalization factor in the denominator $P(d)=\int P(d|g,\omega_{\rm r}) P(g,\omega_{\rm r})\, dg\, d\omega_{\rm r}$ corresponds to average of the likelihood over
all possible values of $(g,\omega_{\rm r})$. One can then apply Bayes' theorem to update the posterior probability distribution $P(g,\omega_{\rm r}|d)$ for the unknown parameters $g$ and $\omega_{\rm r}$ after the measurement. The posterior can then be interpreted as the prior for the next measurement which makes iterative learning of the parameters possible [Fig.~\ref{fig:fig1}(a)]
\begin{equation}
P(g,\omega_{\rm r}|d_{n+1})=\frac{P(d_{n+1}|g,\omega_{\rm r})P(g,\omega_{\rm r}|d_n)}{\int P(d_{n+1}|g,\omega_{\rm r}) P(g,\omega_{\rm r}|d_n) dg d\omega_{\rm r}}.
\label{eq:iter_bayes}
\end{equation}
Once a sufficient amount of data has been collected, {\it e.g.}, after $N_{\rm u}$ updates of the measurement setting, the estimate $(\tilde{g},\tilde{\omega}_{{\rm r}})$ is obtained from the mean over the posterior 
\begin{align}
&\tilde{g}=\int gP(g,\omega_{\rm r}|D) dg d\omega_{\rm r},\ \tilde{\omega}_{\rm r}=\int \omega_{\rm r}P(g,\omega_{\rm r}|D) dg d\omega_{\rm r},\nonumber\\
&D=\{d_1,\ldots,d_{N_{\rm u}} \}.
\end{align}
For numerical implementation of the inference scheme outlined above we adopt sequential Monte Carlo method \cite{west93,gordon93,liu01} that has, due to its computational efficiency, recently gained popularity in the context of quantum tomography \cite{huszar12,granade12,kravtsov13,ferrie13,stenberg14,wiebe14a,wiebe14b,stenberg15}.

\section{Manually optimized policy}
Bayesian inference scheme above, however, does not {\it per se} fix the way in which the measurement settings should be chosen.
In \cite{stenberg14}, the policy that we refer here as $\mathcal{P}_{\rm man}$, was manually constructed. For 
$\mathcal{P}_{\rm man}$, the $M$th measurement setting $(\omega_{\rm q},t)$ is chosen according to the rules
\begin{eqnarray}
t&=&\begin{cases}
\frac{ar_1}{\sigma_{g}}\quad &{\rm if}\ M \le M_0\\
\frac{|a+bz|}{\sigma_{g}}\quad &{\rm if}\  M > M_0
\end{cases},\nonumber\\
\omega_{\rm q}&=&\begin{cases}
\mu_\omega+\left(r_1-\frac{1}{2}\right)\mu_g\quad &{\rm if}\ M \le M_0\\
\mu_\omega+c\left(r_2-\frac{1}{2}\right)\sigma_{\omega}\quad &{\rm if} \ M > M_0.
\end{cases}
\label{eq:manpolicy}
\end{eqnarray}
Here, $\mu_g$ ($\mu_\omega$) is the mean and $\sigma_g$ ($\sigma_\omega$) the standard deviation of $g$ ($\omega_{\rm r}$) calculated from the prior. Moreover, $z$ is a standard normal deviate while $r_{1,2}$ are uniform random variables on the interval $[0,1]$. The numerical constants $a$, $b$, $c$, and $M_0$ were in \cite{stenberg14} optimized manually and set to $a=1.57$, $b=0.518$, $c=3.0$, and $M_0=15$. Note that incorporating the current knowledge about $\omega_{\rm r}$ in $\mu_\omega$ makes it possible to focus the measurements on the relevant frequency range which would not be possible with measurement settings chosen prior data collection. Ideally one would expect that the best way to do this is to choose measurement settings such that $\omega_q-\mu_\omega\sim\sigma_\omega$ [the fourth line in Eq.~(\ref{eq:manpolicy})], reflecting typical parameter uncertainty in the posterior. We find, however, that in order to make the policy more robust against possible numerical error, it is advantageous to also perform measurements such that $\omega_q-\mu_\omega\sim\mu_g$ [the third line in Eq.~(\ref{eq:manpolicy}); cf. also Eq.~(\ref{eq:w_policy2})]. Note that $\mu_g$ determines the expected width of the likelihood function on frequency axis and is thus another relevant frequency scale. 

While the policy of Eq.~(\ref{eq:manpolicy}) performs well in estimating $g$ with longer relaxation times, it is less satisfactory in estimating $\omega_{\rm r}$ when the excitation oscillating between the qubit and the resonator undergoes less than a few tens of Rabi cycles before relaxation takes place. Because estimating $\omega_{\rm r}$ is often of greater significance, {\it e.g.}, in characterizing spurious TLSs, and since the decoherence times of TLSs can vary on a wide range, from a few nanoseconds to microseconds
\cite{shalibo10,palomaki10,lisenfeld10}, new policies are required to characterize them efficiently.

\section{Particle swarm optimization}
In principle, the optimal values of $a$, $b$, $c$, and $M_0$ in Eq.~(\ref{eq:manpolicy}) depend,  {\it e.g.}, on the relaxation time $T_1$, on the possible error rate in the readout of the measurement, as well as on the degree of the initial uncertainty of the estimated parameters. However, since optimizing the coefficients $a$, $b$, $c$, and $M_0$ manually for such different conditions would already be inhibitively time consuming, we need a more automatized way to do it. We adopt the particle swarm optimization (PSO) algorithm \cite{kennedy95,PSObook} that for quantum measurements has previously been applied to phase estimation \cite{hentschel10,hentschel11}. PSO tends to outperform \cite{hayes14} other methods known in the literature such as simulated annealing, genetic algorithms, and routines built in mathematical software. As PSO is able to improve policies based on past experience, our work represents machine learning in the quantum domain. Machine learning literature also includes, {\it e.g.}, algorithms for classification of data which have recently been employed for discriminating quantum measurement trajectories and improving readout \cite{magesan15}.

A candidate policy, which we first parametrize through $a$, $b$, $c$, and $M_0$, can be represented as a \textit{particle} position ${\boldsymbol p}^{(i)}=(p_1,\ldots,p_{N_d})$ in an $N_d$-dimensional policy space (here $N_d=4$). Note that here, the meaning of a ``particle'' is different from that in the context of sequential Monte Carlo method. The cyclical operation of the PSO algorithm is illustrated in Fig.~\ref{fig:fig1}(b). In the beginning, we set $N_{PSO}= 60$ particles in random positions picked from a uniform probability distribution in a
region that we are confident to include the optimal position. For each particle, an approximation $\tilde{S}({\boldsymbol p}^{(i)})$ of the fitness function  $S({\boldsymbol p}^{(i)})$ is computed through $K$ trial runs with random true parameter values. We set $S$ to be the negative median squared error of $\tilde{\omega}_{{\rm r}}$ after $N_{\rm u}=200$ updates of the measurement setting. (The negative sign here is a convention to ensure that the best policies are those who maximize $S$.) For these policies, the measurement setting is updated after each shot, and $\tilde{S}$ is thus evaluated after 200 measurement shots. But in Section VI we also consider repeating measurements $M_{\rm r}$ times at the same setting, in which case we evaluate $\tilde{S}$  after $M_{\rm r}N_{\rm u}$ measurement shots.
 
To calculate the fitness function, we have applied the policy to ensembles of $K= 2000$ simulated samples with randomly chosen parameters $(g_0,\omega_{{\rm r},0})$. Here the subscript $0$ denotes a specific fixed true value, in contrast to the symbols naming a quantity. The values $g_0$ and $\omega_{{\rm r},0}$ are chosen from uniform probability distributions with mean values $\mu_{g,0}$, $\mu_{\omega,0}$ and standard deviations $\sigma_{g,0}$, $\sigma_{\omega,0}$, respectively. Note that $\mu_{g,0}$, $\mu_{\omega,0}$, $\sigma_{g,0}$, and $\sigma_{\omega,0}$, unlike $\mu_{g}$, $\mu_{\omega}$, $\sigma_{g}$, and $\sigma_{\omega}$ in Eq.~(\ref{eq:manpolicy}), do not change between the iterative cycles of the parameter estimation [Fig.~\ref{fig:fig1}(a)]. For each sample we have chosen the initial prior of the $(g,\omega_{\rm r})$ estimate to coincide with the probability distribution from which the true values $(g_0,\omega_{{\rm r},0})$ are randomly picked. This is a natural choice since the latter probability distribution encodes everything that is known about the quantities before data collection. Unless specified otherwise we, have chosen $\sigma_{g,0}=0.25\mu_{g,0}$, $\sigma_{\omega,0}=2\mu_{g,0}$. For each figure below, we calculate the median of the squared error from 10 000 simulated samples.

For each particle, its personal-best position until now, ${\boldsymbol p}^{(i)}_{\rm p}$, is saved. To reduce statistical error, fitness function $\tilde{S}({\boldsymbol p}^{(i)}_{\rm p})$ is resampled at these points over another $K$ trials. The value of the fitness function is then set to the arithmetic average over all its evaluations $\bar{S}({\boldsymbol p}^{(i)}_{\rm p})$. If $\tilde{S}({\boldsymbol p}^{(i)})>\bar{S}({\boldsymbol p}^{(i)}_{\rm p})$ the personal-best position is updated to ${\boldsymbol p}^{(i)}_{\rm p}={\boldsymbol p}^{(i)}$.

The essence of the PSO algorithm is that the particles exchange information with their neighbors. We adopt a fixed ring topology \cite{PSObook,zhao13}, where particles are ordered according to their label $i$, independently of their positions in the policy space, so that each particle has two neighbors, one on each side (the $N_{PSO}$th and the $1st$ particle are neighbors). Each particle keeps a record of the best position ${\boldsymbol p}^{(i)}_{\rm n}$ in its neighborhood (including the particle itself). 

At the $j$th iteration step, the particles obey the equation
\begin{equation}
\boldsymbol{p}_{(j+1)}^{(i)} \leftarrow \boldsymbol{p}_{(j)}^{(i)}+\boldsymbol{v}_{(j)}^{(i)},
\label{eq:p_change}
\end{equation}
where the velocity $\boldsymbol{v}_{(j)}^{(i)}$ is updated through \cite{shi98,zahedinejad14}
\begin{equation}
\boldsymbol{v}_{(j+1)}^{(i)} \leftarrow w\boldsymbol{v}_{(j)}^{(i)}+\beta_1\xi_1(\boldsymbol{p}_{{\rm p},(j)}^{(i)}-\boldsymbol{p}_{(j)}^{(i)})+\beta_2\xi_2(\boldsymbol{p}_{{\rm n},(j)}^{(i)}-\boldsymbol{p}_{(j)}^{(i)}).
\label{eq:v_change}
\end{equation}
The constants $\beta_1$ and $\beta_2$ determine the attraction of a particle at $\boldsymbol{p}_{(j)}^{(i)}$ to $\boldsymbol{p}_{{\rm p},(j)}^{(i)}$ and $\boldsymbol{p}_{{\rm n},(j)}^{(i)}$, respectively, while $\xi_1$ and $\xi_2$ are uniform random variables on the interval $[0,1]$. The constant $w$, referred as inertia weight, introduces a friction-like effect. We choose the parameter values $\beta_1=0.5$,  $\beta_2=1.0$, and $w=0.7$. To make the convergence towards global, rather than local, optimum more likely, we bound the absolute value of each component of $\boldsymbol{v}_{(j)}^{(i)}$ by a maximum value. We iterate Eqs.~(\ref{eq:p_change}) and (\ref{eq:v_change}) until we see the fitness function to saturate.

\begin{figure}[h!]
\includegraphics[width=0.55\textwidth]{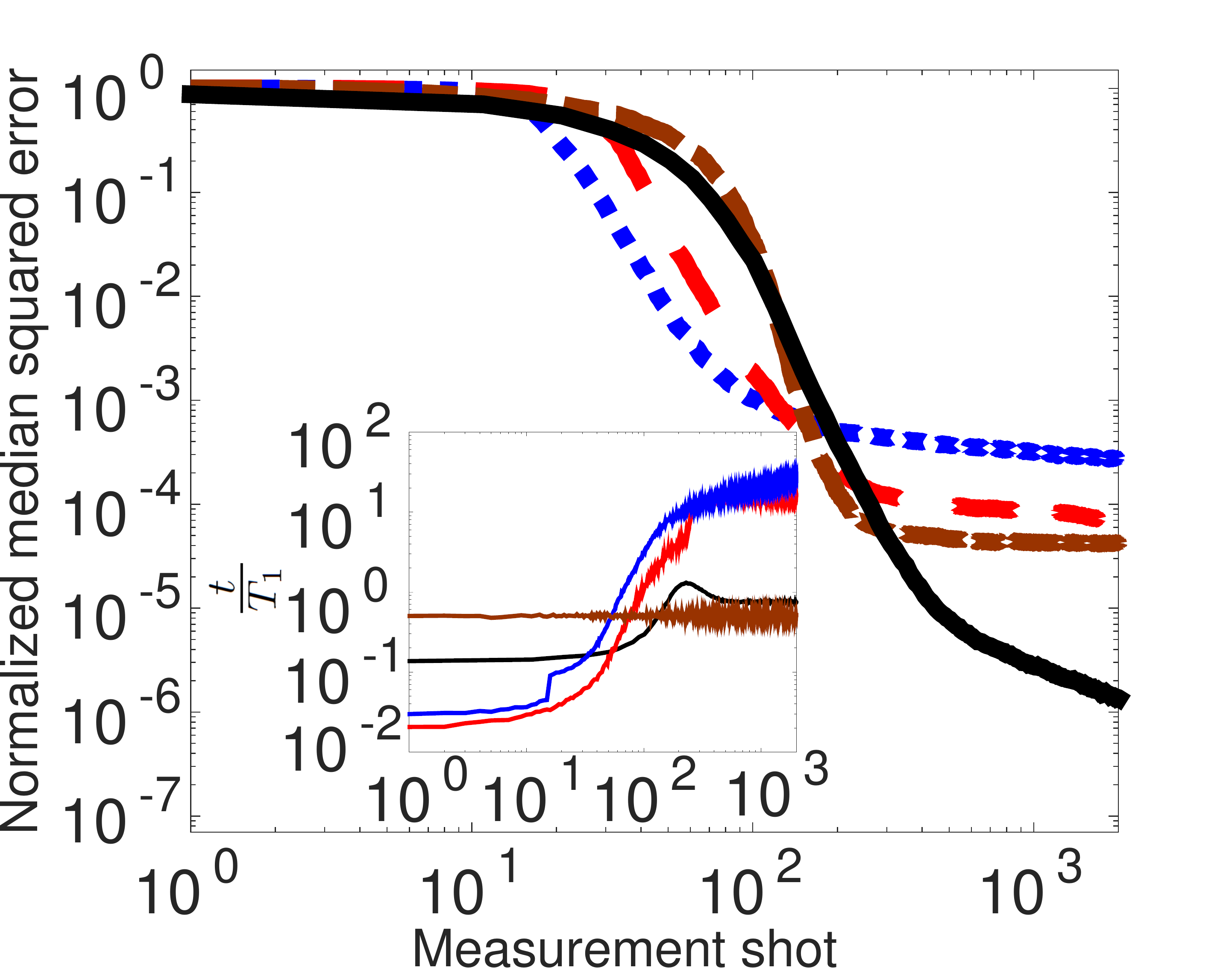}
\caption{(Color online) Normalized median squared error of $\tilde{\omega}_{{\rm r}}$ calculated from an ensemble of simulated samples (see text) with the normalized relaxation time $T_1\mu_{g,0}=20\pi$. The curves correspond to the policies $\mathcal{P}_{\rm man}$ (blue dotted),  $\tilde{\mathcal{P}}_{\rm mach}^{(20)}$ (red dashed), $\mathcal{P}_{\rm rand}$ (brown dash-dotted), and $\mathcal{P}_{\rm mach,u}^{(20)}$ (black solid). See the main text for the description of the policies. Inset: Normalized median waiting times for the same policies (colors as in the main figure).}
\label{fig:errcompare}
\end{figure} 

\section{Policies optimized through PSO}
We have optimized the coefficients $a$, $b$, $c$, and $M_0$ of Eq.~(\ref{eq:manpolicy}) through PSO in the presence of relaxation to minimize the error of the estimate. For all the figures below, we normalize the error so that for the initial prior, the normalized median squared error calculated from the simulated ensemble equals unity. Figure~\ref{fig:errcompare} exhibits the normalized median squared error of $\tilde{\omega}_{{\rm r}}$ for the policy that we refer as $\tilde{\mathcal{P}}_{\rm mach}^{(20)}$, obtained for $T_1\mu_{g,0}=20\pi$ as described above. While $\tilde{\mathcal{P}}_{\rm mach}^{(20)}$ improves the accuracy of the estimate in the beginning, after some hundreds of measurement shots the median error saturates, which indicates that measurements are not informative. Indeed, after $O(10^2)$ measurement shots, $\tilde{\mathcal{P}}_{\rm mach}^{(20)}$ yields waiting times $t$ that are so long compared to the relaxation time $T_1$ that the amplitude of the Rabi-oscillations has mostly decayed (see the inset of Fig.~\ref{fig:errcompare}). Since Eq.~(\ref{eq:manpolicy}) connects $t$ to the precision of the estimate, it is, at the first site, surprising that $t$ can become longer but the estimates do not improve. The explanation is that the posteriors do become narrower but converge to a wrong value. Also for the simple policy $\mathcal{P}_{\rm rand}$ (the brown curve in Fig.~\ref{fig:errcompare}) the median error saturates after having reached a floor whose level we attribute to balanced competition between information obtained through measurements and accumulated numerical error. For $\mathcal{P}_{\rm rand}$, $t$ is random on a uniform probability density on the time interval  $[0,T_1]$, while $\omega_{\rm q}$ obeys Eq.~(\ref{eq:manpolicy}) adaptively. Even though ideally one would not expect it, with uninformative measurements, due to loss of numerical accuracy, Bayesian inference schemes quite commonly lead to accuracy saturating at a certain level. We emphasize that this is characteristic (but not unavoidable) in the presence of strong relaxation where signal is rapidly lost. In the absence of relaxation, the policies $\mathcal{P}_{\rm man}$ and $\tilde{\mathcal{P}}_{\rm mach}^{(20)}$ achieve errors far below the current levels. Characterizing decohering quantum systems thus requires special effort.

To find more efficient policies, we have modified the policy space for the search of the optimal policy. We introduce a maximum $t_{\rm max}$ for the waiting times such that we always have $t<t_{\rm max}$ once a certain level of accuracy has been reached. Moreover, to overcome the numerical error that accumulates in updating the posterior (see \cite{wiebe14a}
for more details), we repeat the measurement at the same setting $M_{\rm r}=10$ times before updating the posterior. Hence, for a single measurement setting, in Eqs.~({\ref{eq:bayes_theorem}) and (\ref{eq:iter_bayes}) we have for the possible number of measurement shots that detect the qubit in its ground state  $d=0,1,\ldots,10$. For different measurement settings, we keep a record on the number of times $C$ for which $d$ has exceeded a threshold value $D_{\rm th}$. In the following, the event of obtaining $d>D_{\rm th}$ for a certain measurement setting plays a similar role as detecting a `click' for a single measurement shot. Note that as $\frac{d}{M_{\rm r}}$ gives a crude estimate for the probability for the qubit to be in the ground state, the event of obtaining $d>D_{\rm th}$ implies this estimate exceeding a threshold value $P_{\rm th}=\frac{D_{\rm th}}{M_{\rm r}}$. For the candidate policies studied below, we define the fitness function $S$ to be taken after $M_{\rm r}N_{\rm u} = 10\times 200=2000$ measurement shots.
\begin{figure}[h!]
\includegraphics[width=0.5\textwidth]{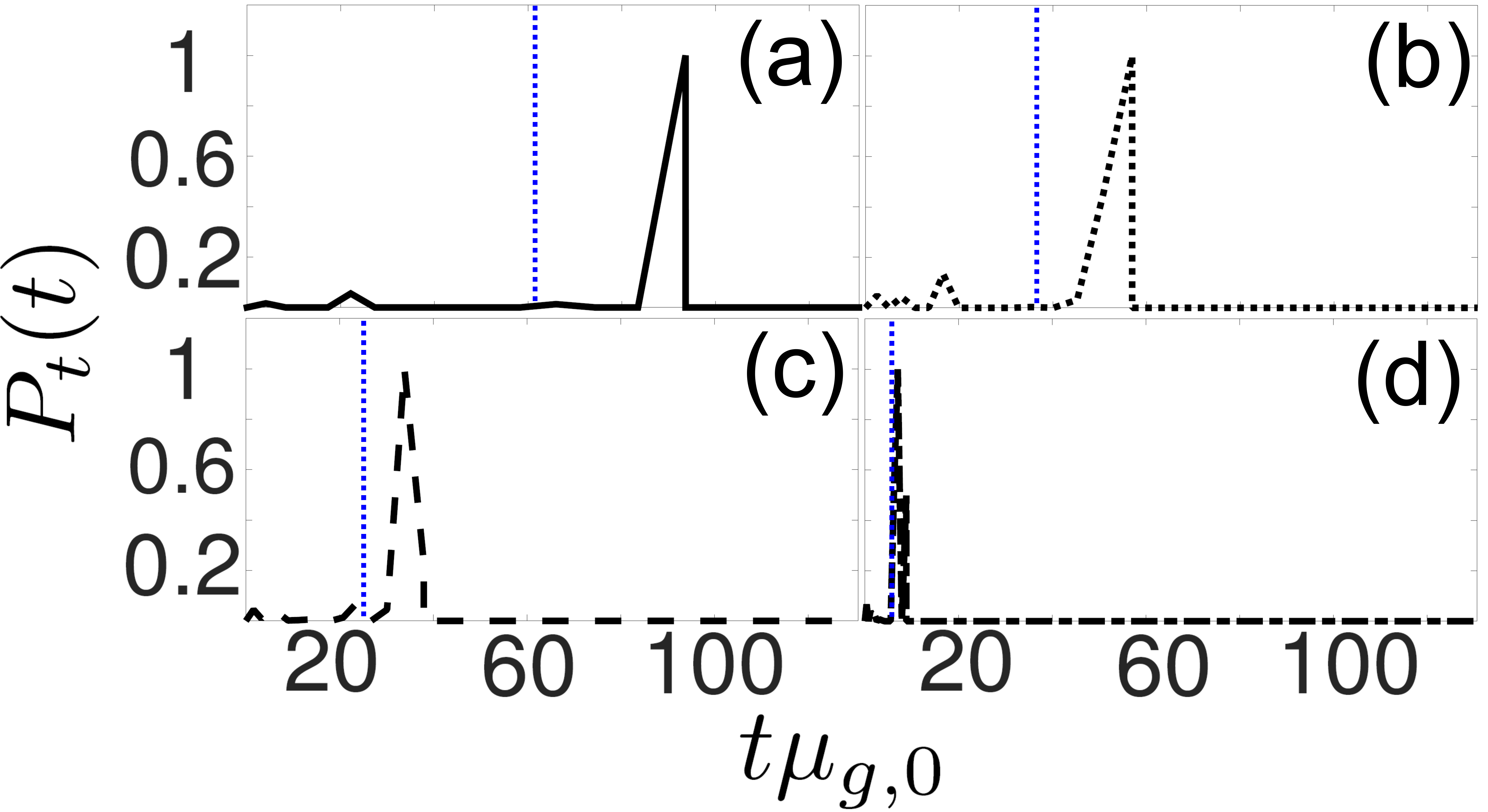}
\includegraphics[width=0.5\textwidth]{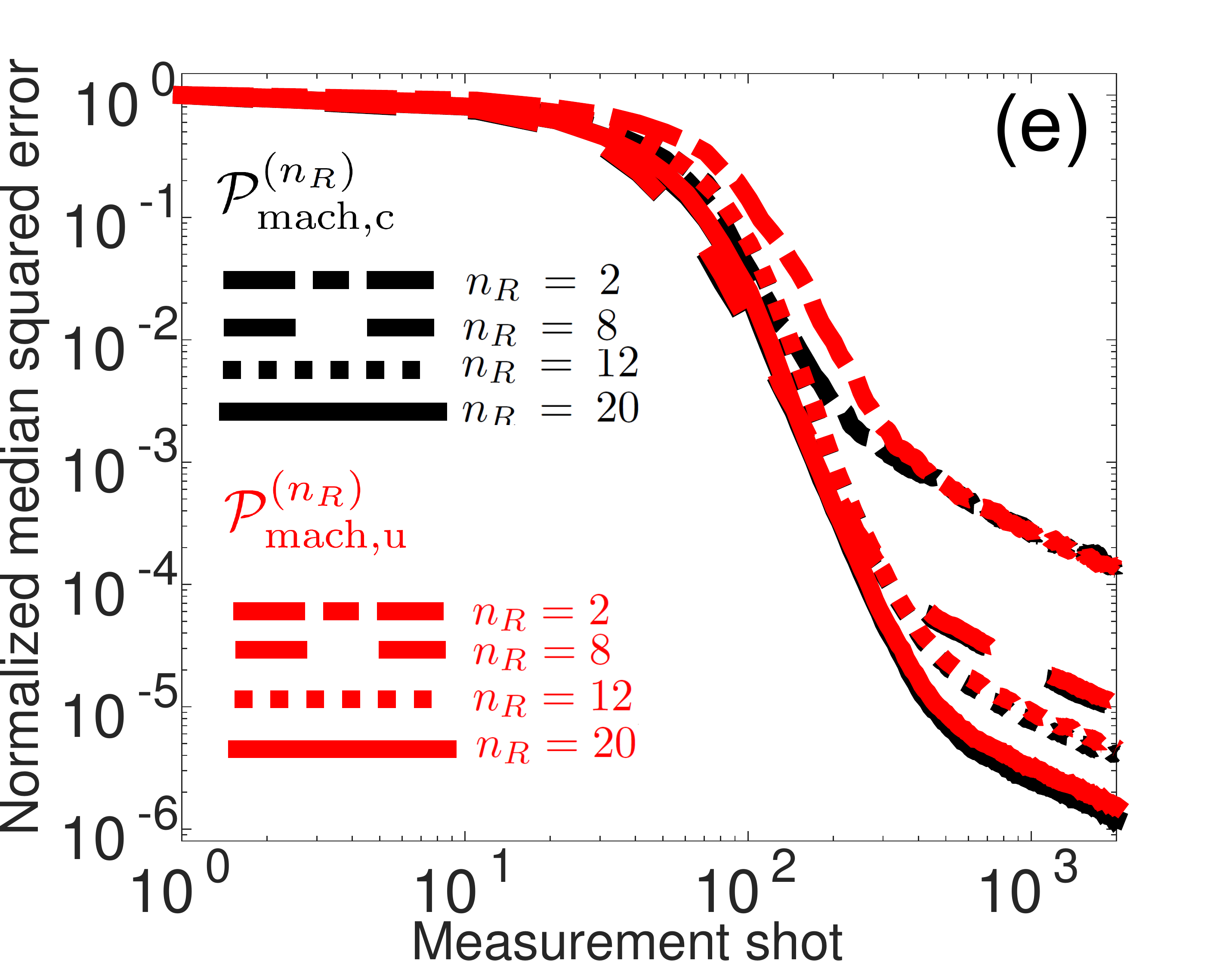}
\caption{(Color online) Optimal shape for the probability density $P_t(t)$ (see text).
The normalized relaxation times are $T_1\mu_{g,0}=n_{\rm R}\pi$ with the number of vacuum Rabi cycles $n_{\rm R}$ equal to (a) 20, (b) 12, (c) 8, and  (d) 2; the panels exhibit $P_t(t)$
obtained for the policies $\mathcal{P}_{\rm mach,c}^{(n_{\rm R})}$. The dashed vertical lines mark the normalized relaxation times $T_1\mu_{g,0}$. Panel (e) exhibits the normalized median squared errors for $\tilde{\omega}_{{\rm r}}$ calculated for ensembles of simulated samples. In (e), the parameter values correspond to the panels (a)-(d): $n_{\rm R}=20$ (solid), $n_{\rm R}=12$ (dotted), $n_{\rm R}=8$ (dashed), and $n_{\rm R}=2$ (dash-dotted). Black curves are for the policies $\mathcal{P}_{\rm mach,c}^{(n_{\rm R})}$, red curves for $\mathcal{P}_{\rm mach,u}^{(n_{\rm R})}$ that yield near-optimal results.} 
\label{fig:fig3}
\end{figure} 

We put forward policies $\mathcal{P}_{\rm mach,u}^{(n_{\rm R})}$ (the meaning of the sub and superscripts will become clear below) where the measurement setting $(\omega_{\rm q},t)$ depends on the accuracy that has been reached or the relative magnitude of $\sigma_g$ and $\frac{1}{t_{\rm max}}$, as well as on $C$. First, with $C=0$, we pick $\omega_{\rm q}$ randomly from a probability distribution defined by the current prior of $\omega_{\rm r}$ 
\begin{equation}
 P_{\omega_{\rm q}}=P(\omega_{\omega_{\rm q}}|D),\quad D=\{d_1,\ldots, d_M\},
 \label{eq:P_omega_q}
\end{equation}
which makes probing too similar frequencies twice unlikely, whereas $t$ obeys the first line in the rule
\begin{equation}
t=\begin{cases}
\frac{ar_1}{\sigma_{g}}\quad &{\rm if}\ C \le C_0\\
\frac{|d+bz|}{\sigma_{g}}\quad &{\rm if}\  C >  C_0.
\end{cases}
\label{eq:t_policy2}
\end{equation}
Second, if $C>0$ and $\sigma_g>\frac{1}{t_{\rm max}}$, we choose $\omega_{\rm q}$ using the equations
\begin{equation}
\omega_{\rm q}=\begin{cases}
\mu_\omega+f\left(r_1-\frac{1}{2}\right)\mu_g\quad &{\rm if}\ 1 \le C \le C_0\\
\mu_\omega+g\left(r_2-\frac{1}{2}\right)\sigma_{\omega}\quad &{\rm if} \ C > C_0.
\end{cases}
\label{eq:w_policy2}
\end{equation}
and $t$ by Eq.~(\ref{eq:t_policy2}).
Above, $a$, $b$, $d$, $f$, $g$ $\in {\mathbb R}$ and  $D_{\rm th},C_0 \in {\mathbb N}$, as well as $t_{\rm max}\mu_{g,0}\in {\mathbb R}$ with a fixed $\mu_{g,0}$, are parameters that we optimize through PSO. The random variables $r_{1,2}$, $z$ have the same meaning as in Eq.~(\ref{eq:manpolicy}). In the third case with $C>0$ and $\sigma_g\le \frac{1}{t_{\rm max}}$, we update $\omega_{\rm q}$ as in Eq.~(\ref{eq:w_policy2}), but choose $t$ randomly from probability distribution $P_t(t)$ that is uniform on a time interval $[0,t_{\rm max}]$. 
The policy $\mathcal{P}_{\rm mach,u}^{(n_{\rm R})}$, with $n_{\rm R}=20$, significantly outperforms the other policies mentioned above as illustrated in Fig.~\ref{fig:errcompare}, and the corresponding median error does not saturate on the considered interval of measurement shots. As shown in the inset, in the beginning of the experiment, the waiting times used by $\mathcal{P}_{\rm mach,u}^{(20)}$ are longer than those of  $\mathcal{P}_{\rm man}$ and $\tilde{\mathcal{P}}_{\rm mach}^{(20)}$, and with a large number of measurement shots, the exponential increase of $t$ is limited by $t_{\rm max}$.
 
We find that uniform $P_t(t)$ yields results close to those for the optimal shape of $P_t(t)$. To demonstrate this, we study policies $\mathcal{P}_{\rm mach,c}^{(n_{\rm R})}$ where we
allow $P_t(t)$ {\it a priori} to be a more general function, while keeping the parameters characterizing $\mathcal{P}_{\rm mach,u}^{(n_{\rm R})}$ fixed at their optimal values. We set $N_t=16$ time points $0=t_1<t_2,\ldots<t_{N_t}=t_{\rm max}'$ such that their density is inversely proportional to the envelope of the exponentially decaying Rabi oscillations $\sim\exp(-t/T_1)$. We then search for an optimal $P_t(t)$ that is a continuous linear function between these time points $t_i$. The values $P_t(t_i)$ determine $P_t(t)$ uniquely and we optimize them through PSO. 

 Figures~\ref{fig:fig3}(a)-(d) exhibit $P_{t}(t)$ optimized for different relaxation times, for $T_1\mu_{g,0}=n_{\rm R}\pi$ with $n_{\rm R}=20,\ 12,\ 8,$ and $2$, respectively. Note that $n_{\rm R}=\frac{T_1\mu_{g,0}}{\pi}$ is the number of Rabi cycles that an excitation oscillating between the qubit and the resonator undergoes in time $T_1$ when the coupling strength is $\mu_{g,0}$ (for details, see \cite{stenberg14} and references therein). Our PSO calculations indicate that optimally $P_{t}(t)$ is peaked around the value that lies on the time interval 
 $[1.31T_1,1.49T_1]$. However, we find that the median error is not very sensitive to the shape of $P_t(t)$, as long as the mean of the distribution is of the correct order of magnitude, and uniform $P_t(t)$ can give near-optimal results. Optimizing $\mathcal{P}_{\rm mach,u}^{(n_{\rm R})}$ corresponds to exploring an 8-dimensional policy space. The parameter values determining the best policy found within this space are presented in Appendix in Table I. For $t_{\rm max}$, we find advantageous to set it in the region $[1.37T_1, 1.55T_1]$.

In Fig.~\ref{fig:fig3}(e) we plot the normalized median squared error of $\tilde{\omega}_{{\rm r}}$ for the relaxation times $T_1\mu_{g,0}=20\pi$ (solid), $T_1\mu_{g,0}=12\pi$ (dotted),  $T_1\mu_{g,0}=8\pi$ (dashed),  and $T_1\mu_{g,0}=2\pi$ (dash-dotted), obtained for the corresponding optimized policies $\mathcal{P}_{\rm mach,u}^{(n_{\rm R})}$ (black) and $\mathcal{P}_{\rm mach,c}^{(n_{\rm R})}$ (red). We find that $\mathcal{P}_{\rm mach,u}^{(n_{\rm R})}$ and $\mathcal{P}_{\rm mach,c}^{(n_{\rm R})}$ yield approximately the same error and uniform $P_t(t)$ thus yields near-optimal results. In the following we therefore concern ourselves on the policies of the type $\mathcal{P}_{\rm mach,u}^{(n_{\rm R})}$.

\begin{figure}
\includegraphics[width=0.5\textwidth]{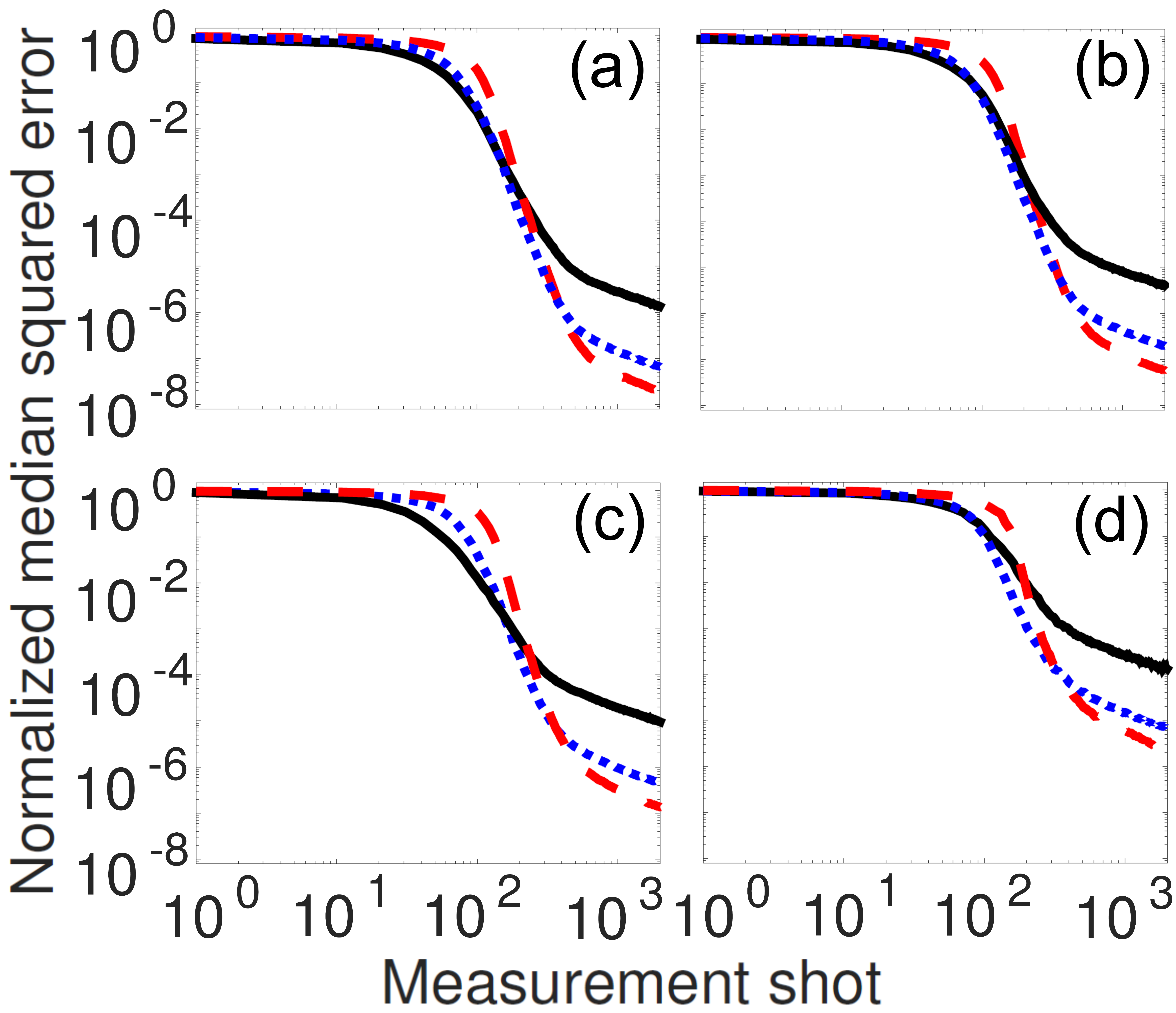}
\caption{(Color online) The influence of the initial parameter uncertainty. Normalized median squared error for ensembles of simulated samples, with standard deviations 
$\sigma_{\omega,0}$ equal to $2\mu_{g,0}$ (black solid), $10\mu_{g,0}$ (blue dotted), and $20\mu_{g,0}$ (red dashed). Different panels correspond to normalized relaxation times  
$T_1\mu_{g,0}=n_{\rm R}\pi$, with $n_{\rm R}$ equal to (a) 20, (b) 12, (c) 8, and  (d) 2. All the curves have been calculated through policies $\mathcal{P}_{\rm mach,u}^{(n_{\rm R},\sigma)}$ optimized for different $T_1\mu_{g,0}$ and $\sigma_{\omega,0}$.}
\label{fig:pr}
\end{figure} 
\section{Extreme initial uncertainty}
To optimize the performance of the policies employing a Bayesian inference scheme, they also have to be adjusted for the initial parameter uncertainty and the initial prior. 
The performance for different initial priors is illustrated in Figs.~\ref{fig:pr}. We consider normalized relaxation times $T_1\mu_{g,0}=n_{\rm R}\pi$, with $n_{\rm R}$ equal to (a) 20, (b) 12, (c) 8, and (d) 2. For each relaxation time, we study different standard deviations for the uniform probability distribution through which the true values $(g_0,\omega_{{\rm r},0})$ are randomly chosen. Different curves in Fig.~\ref{fig:pr} are for $\sigma_{\omega,0}=2 \mu_{g,0}$ (black solid), $\sigma_{\omega,0}=10 \mu_{g,0}$ (blue dotted), and $\sigma_{\omega,0}=20 \mu_{g,0}$ (red dashed), calculated through policies that we denote by $\mathcal{P}_{\rm mach,u}^{(n_{\rm R},2)}=\mathcal{P}_{\rm mach,u}^{(n_{\rm R})}$, $\mathcal{P}_{\rm mach,u}^{(n_{\rm R},10)}$, and $\mathcal{P}_{\rm mach,u}^{(n_{\rm R},20)}$, respectively. Each policy is optimized for the corresponding $T_1$ and $\sigma_{\omega,0}$. Throughout Fig.~\ref{fig:pr} we have  $\sigma_{g,0}=0.25\mu_{g,0}$. The parameters determining the policies are listed in Appendix in Table I.  Our policies are robust also for extreme initial uncertainty in $\omega_{\rm r}$. With a large uncertainty in $\omega_{\rm r}$, the qubit frequency $\omega_{\rm q}$ follows the rule (\ref{eq:P_omega_q}), but with the first detection of $d>D_{\rm th}$ the uncertainty decreases to $O(\mu_g)$. This is why median error exhibits an increasingly steep drop with increasing initial uncertainty $\sigma_{\omega,0}$.

\begin{figure}[h!]
\includegraphics[width=0.5\textwidth]{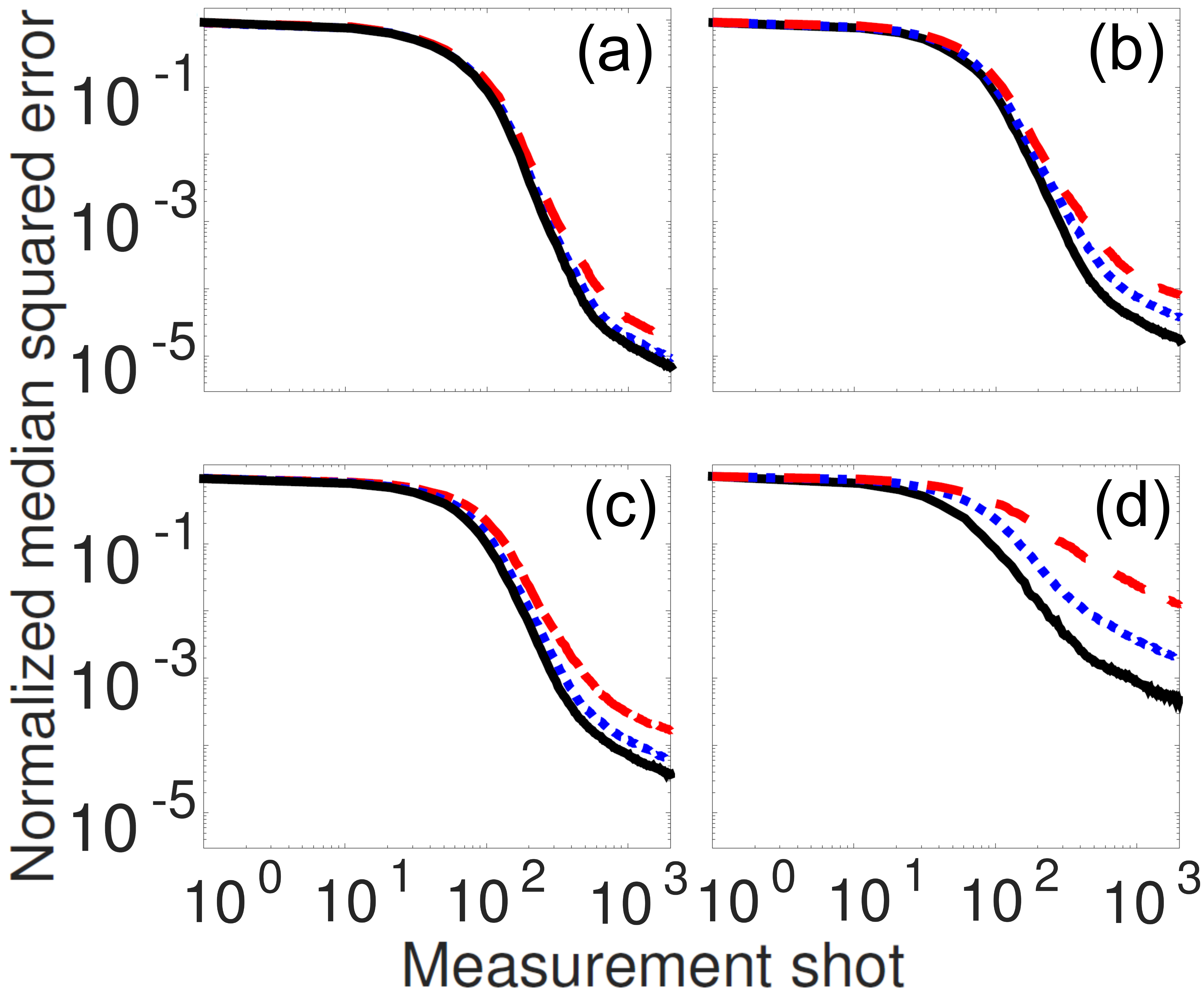}
\caption{(Color online) Performance with imperfections in the readout, with the error probability $P_{\rm e}=0.1$. Normalized median squared error calculated for ensembles of simulated samples for policies $\mathcal{P}_{\rm mach,u}^{(n_{\rm R}),re}$ with presumed $n_{\rm R}$ equal to (a) 20, (b) 12, (c) 8, and (d) 2. The true normalized relaxation times $T_1\mu_{g,0}$ are (a) $20\pi$ (black solid), $16\pi$ (blue dotted), $12\pi$ (red dashed), (b) $12\pi$ (black solid), $8\pi$ (blue dotted), $6\pi$ (red dashed), (c) $8\pi$ (black solid), $6\pi$ (blue dotted), $4\pi$ (red dashed), and (d) $2\pi$ (black solid), $\pi$ (blue dotted), $\frac{\pi}{2}$ (red dashed). }
\label{fig:re}
\end{figure} 
\section{Experimental imperfections}
To account for possible presence of experimental imperfections, we assume a finite probability $P_{\rm e}$ for a readout error or the probability of misidentifying a ground state as  an excited state and vice versa. As a conservative estimate, we set $P_{\rm e}=0.1$. In Figs.~\ref{fig:re}(a)-(d) the black curves exhibit the normalized median squared error of $\tilde{\omega}_{{\rm r}}$ for the policies $\mathcal{P}_{\rm mach,u}^{(20),{\rm re}}$, $\mathcal{P}_{\rm mach,u}^{(12),{\rm re}}$, $\mathcal{P}_{\rm mach,u}^{(8),{\rm re}}$, and $\mathcal{P}_{\rm mach,u}^{(2),{\rm re}}$, optimized for the normalized relaxation times $T_1\mu_{g,0}=20\pi$, $12\pi$, $8\pi$, and $2\pi$, respectively. As $T_1$ is not typically known precisely and can also vary as a function of time \cite{chang13}, we also test $\mathcal{P}_{\rm mach,u}^{(n_{\rm R}),{\rm re}}$ for the situations where the true relaxation time differs from that presumed by the policy. We find that if the true relaxation time is longer than that presumed by the policy, this nevertheless leads to more accurate estimates (not shown in the figures). In Fig.~\ref{fig:re} we study a more challenging situation where the true relaxation time is shorter. We consider (a) $\mathcal{P}_{\rm mach,u}^{(20),{\rm re}}$ with the true $T_1\mu_{g,0}=20\pi, 16\pi, 12\pi$, (b) $\mathcal{P}_{\rm mach,u}^{(12),{\rm re}}$ with the true $T_1\mu_{g,0}=12\pi, 8\pi, 6\pi$, (c) $\mathcal{P}_{\rm mach,u}^{(8),{\rm re}}$ with the true $T_1\mu_{g,0}=8\pi, 6\pi, 4\pi$, and (d) $\mathcal{P}_{\rm mach,u}^{(2),{\rm re}}$ with the true $T_1\mu_{g,0}=2\pi, \pi, \frac{\pi}{2}$. 
When the vast majority of waiting times exceed the true $T_1$, this leads to loss of signal which can limit the accuracy of an estimate. But we find improving estimates even when the true $T_1$ is only $\sim 50 \%$ of the value presumed by the policy. These results suggest that a more advanced method where $T_1$ is treated as another unknown parameter would be robust against relatively large amount of error in $T_1$.

\section{Conclusion}
In conclusion, we note that decoherence, an ubiquitous source of errors in quantum information processing devices, continues to limit their performance and our ability to characterize them efficiently. For instance, in the system studied in this paper, much of the power of adaptive measurements is based on the idea to employ exponentially increasing, rather than evenly spaced waiting times. In the absence of relaxation, this can exponentially speed up the estimation. However, if the system can only undergo a small number of Rabi oscillations before relaxation takes place, this sets a limit on the longest useful waiting time. Taking into account this limitation, we have used machine learning techniques to make the measurements as informative as possible. We have delivered policies which make system characterization vastly more efficient also in the presence of experimental imperfections.

\section*{ACKNOWLEDGEMENTS}
We acknowledge A. Hentschel, P. Palittapongarnpim, and B. C. Sanders for discussions. This work was supported by the European Union through ScaleQIT. 

\section*{APPENDIX}
Every candidate policy for $\mathcal{P}_{\rm mach,u}^{(n_{\rm R},\sigma)(,{\rm re})}$ can be parametrized as a vector $(a,b,d,f,g,t_{\rm max}\mu_{g,0},D_{\rm th},C_0)\in {\mathbb R}^6\times {\mathbb N}^2$. Similarly, candidate policies for $\mathcal{P}_{\rm mach,c}^{(n_{\rm R},\sigma)(,{\rm re})}$ can be parametrized as a vector $[P_t(t_2),\ldots,P_t(t_{16})]\in {\mathbb R}^{15}$. The parameters determining our best policy $\mathcal{P}_{\rm mach,u}^{(n_{\rm R},\sigma)(,{\rm re})}$ are presented in Table I; for those of $\mathcal{P}_{\rm mach,c}^{(n_{\rm R},\sigma)(,{\rm re})}$, see Figs.~\ref{fig:fig3}(a)-(d).
\begin{table}[h!]
\begin{center} 
\begin{tabular}{c || c | c | c | c | c | c | c | c  }
$\mathcal{P}/$param.& $a$ & $b$ & $d$ & $f$ &  $g$  & $t_{\rm max}\mu_{g,0}$& $D_{\rm th}$&  $C_0$   \\ 
\hline\hline
$\mathcal{P}_{\rm mach,u}^{(2,2)}$    & 3.92 &  5.61&0.94 & 5.04 & 3.47& 9.18 & 6& 190  \\
\hline
$\mathcal{P}_{\rm mach,u}^{(2,2),\mathrm{re}}$   & 1.45  & 3.52 & 3.14&  6.28 & 1.38 &  9.17& 5& 118   \\
\hline
$\mathcal{P}_{\rm mach,u}^{(2,10)}$  & 1.29  & 3.53 & 3.09& 4.44 & 4.90 & 9.17& 9& 28   \\
\hline
$\mathcal{P}_{\rm mach,u}^{(2,20)}$  & 1.04  & 3.41& 3.13 & 5.41 & 3.18 & 9.17& 8& 198   \\
\hline
$\mathcal{P}_{\rm mach,u}^{(8,2)}$ & 2.65  & 2.21& 8.58 & 4.72 & 6.16 & 37.95& 3& 195  \\
\hline
$\mathcal{P}_{\rm mach,u}^{(8,2),\mathrm{re}}$  & 3.88  & 2.16& 0.73 & 4.46 & 1.05 & 34.93& 2&  194   \\
\hline
$\mathcal{P}_{\rm mach,u}^{(8,10)}$ & 3.13 &0.00& 3.56&3.83 & 0.91 & 36.19& 8& 61   \\
\hline
$\mathcal{P}_{\rm mach,u}^{(8,20)}$  & 3.56  & 1.39 & 5.37& 5.03 & 3.21 & 35.02 & 3&  195  \\
\hline
$\mathcal{P}_{\rm mach,u}^{(12,2)}$ & 8.68  & 7.37 & 4.38& 4.14 & 5.34 & 56.17& 5& 121  \\
\hline
$\mathcal{P}_{\rm mach,u}^{(12,2),\mathrm{re}}$  & 3.88  & 6.02 & 2.71& 5.18 & 4.84  & 55.42& 1& 199  \\
\hline
$\mathcal{P}_{\rm mach,u}^{(12,10)}$  & 4.57  & 0.00 & 1.76& 4.74 & 1.48 & 52.78 & 7& 159  \\
\hline
$\mathcal{P}_{\rm mach,u}^{(12,20)}$  & 3.79  & 0.00 & 2.81& 4.57 & 3.79 & 58.43& 7 &  76  \\ 
\hline
$\mathcal{P}_{\rm mach,u}^{(20,2)}$ &  7.49  & 3.11& 1.44 & 4.96 & 5.98 & 94.25& 6&  129  \\
\hline
$\mathcal{P}_{\rm mach,u}^{(20,2),\mathrm{re}}$  & 7.31  & 2.83& 0.00 & 4.73 & 4.57 & 86.08 & 8& 199  \\
\hline
$\mathcal{P}_{\rm mach,u}^{(20,10)}$  & 5.38  & 0.00 & 0.84& 3.88 & 0.33 & 93.62& 9& 199  \\
\hline
$\mathcal{P}_{\rm mach,u}^{(20,20)}$  & 4.02  & 0.06 & 5.87& 4.74 & 0.00 & 97.34& 6& 94   \\
\hline
\end{tabular} 
\caption{Parameters determining optimized policies $\mathcal{P}$. For the meaning of the parameters, see Eqs.~(\ref{eq:t_policy2}) and (\ref{eq:w_policy2}) and the discussion above them.}
\label{table1} 
\end{center} 
\end{table}


\begin{thebibliography}{41}
\expandafter\ifx\csname natexlab\endcsname\relax\def\natexlab#1{#1}\fi
\expandafter\ifx\csname bibnamefont\endcsname\relax
  \def\bibnamefont#1{#1}\fi
\expandafter\ifx\csname bibfnamefont\endcsname\relax
  \def\bibfnamefont#1{#1}\fi
\expandafter\ifx\csname citenamefont\endcsname\relax
  \def\citenamefont#1{#1}\fi
\expandafter\ifx\csname url\endcsname\relax
  \def\url#1{\texttt{#1}}\fi
\expandafter\ifx\csname urlprefix\endcsname\relax\def\urlprefix{URL }\fi
\providecommand{\bibinfo}[2]{#2}
\providecommand{\eprint}[2][]{\url{#2}}

\bibitem[{\citenamefont{{Y. Nakamura} et~al.}(1999)\citenamefont{{Y. Nakamura},
  {Yu. A. Pashkin}, and {J. S. Tsai}}}]{nakamura99}
\bibinfo{author}{\bibnamefont{{Y. Nakamura}}},
  \bibinfo{author}{\bibnamefont{{Yu. A. Pashkin}}}, \bibnamefont{and}
  \bibinfo{author}{\bibnamefont{{J. S. Tsai}}}, \bibinfo{journal}{Nature
  (London)} \textbf{\bibinfo{volume}{398}}, \bibinfo{pages}{786}
  (\bibinfo{year}{1999}).

\bibitem[{\citenamefont{{M. H. Devoret} and {R. J.
  Schoelkopf}}(2013)}]{devoret13}
\bibinfo{author}{\bibnamefont{{M. H. Devoret}}} \bibnamefont{and}
  \bibinfo{author}{\bibnamefont{{R. J. Schoelkopf}}},
  \bibinfo{journal}{Science} \textbf{\bibinfo{volume}{339}},
  \bibinfo{pages}{1169} (\bibinfo{year}{2013}).

\bibitem[{\citenamefont{{A. G. Fowler} et~al.}(2012)\citenamefont{{A. G.
  Fowler}, {M. Mariantoni}, {J. M. Martinis}, and {A. N. Cleland}}}]{fowler12}
\bibinfo{author}{\bibnamefont{{A. G. Fowler}}},
  \bibinfo{author}{\bibnamefont{{M. Mariantoni}}},
  \bibinfo{author}{\bibnamefont{{J. M. Martinis}}}, \bibnamefont{and}
  \bibinfo{author}{\bibnamefont{{A. N. Cleland}}}, \bibinfo{journal}{Phys. Rev.
  A} \textbf{\bibinfo{volume}{86}}, \bibinfo{pages}{032324}
  (\bibinfo{year}{2012}).

\bibitem[{\citenamefont{{I. Chuang} and {M. Nielsen}}(1997)}]{chuang97}
\bibinfo{author}{\bibnamefont{{I. Chuang}}} \bibnamefont{and}
  \bibinfo{author}{\bibnamefont{{M. Nielsen}}}, \bibinfo{journal}{J. Mod. Opt.}
  \textbf{\bibinfo{volume}{44}}, \bibinfo{pages}{2455} (\bibinfo{year}{1997}).

\bibitem[{\citenamefont{{J. Emerson} et~al.}(2005)\citenamefont{{J. Emerson},
  {R. Alicki}, and {K. Zyczkowski}}}]{emerson05}
\bibinfo{author}{\bibnamefont{{J. Emerson}}}, \bibinfo{author}{\bibnamefont{{R.
  Alicki}}}, \bibnamefont{and} \bibinfo{author}{\bibnamefont{{K. Zyczkowski}}},
  \bibinfo{journal}{J. Opt. B} \textbf{\bibinfo{volume}{7}},
  \bibinfo{pages}{S347} (\bibinfo{year}{2005}).

\bibitem[{\citenamefont{{E. Knill {\it et al.}}}(2008)}]{knill08}
\bibinfo{author}{\bibnamefont{{E. Knill {\it et al.}}}},
  \bibinfo{journal}{Phys. Rev. A} \textbf{\bibinfo{volume}{77}},
  \bibinfo{pages}{012307} (\bibinfo{year}{2008}).

\bibitem[{\citenamefont{{E. Magesan} et~al.}(2011)\citenamefont{{E. Magesan},
  {J. M. Gambetta}, and {J. Emerson}}}]{magesan11}
\bibinfo{author}{\bibnamefont{{E. Magesan}}}, \bibinfo{author}{\bibnamefont{{J.
  M. Gambetta}}}, \bibnamefont{and} \bibinfo{author}{\bibnamefont{{J.
  Emerson}}}, \bibinfo{journal}{Phys. Rev. Lett.}
  \textbf{\bibinfo{volume}{106}}, \bibinfo{pages}{180504}
  (\bibinfo{year}{2011}).

\bibitem[{\citenamefont{{E. Magesan {\it et al.}}}(2012)}]{magesan12}
\bibinfo{author}{\bibnamefont{{E. Magesan {\it et al.}}}},
  \bibinfo{journal}{Phys. Rev. Lett.} \textbf{\bibinfo{volume}{109}},
  \bibinfo{pages}{080505} (\bibinfo{year}{2012}).

\bibitem[{\citenamefont{{D. W. Berry} et~al.}(2009)\citenamefont{{D. W. Berry},
  {B. L. Higgins}, {S. D. Bartlett}, {M. W. Mitchell}, {G. J. Pryde}, and {H.
  M. Wiseman}}}]{berry09}
\bibinfo{author}{\bibnamefont{{D. W. Berry}}},
  \bibinfo{author}{\bibnamefont{{B. L. Higgins}}},
  \bibinfo{author}{\bibnamefont{{S. D. Bartlett}}},
  \bibinfo{author}{\bibnamefont{{M. W. Mitchell}}},
  \bibinfo{author}{\bibnamefont{{G. J. Pryde}}}, \bibnamefont{and}
  \bibinfo{author}{\bibnamefont{{H. M. Wiseman}}}, \bibinfo{journal}{Phys. Rev.
  A} \textbf{\bibinfo{volume}{80}}, \bibinfo{pages}{052114}
  (\bibinfo{year}{2009}).

\bibitem[{\citenamefont{{A. Sergeevich} et~al.}(2011)\citenamefont{{A.
  Sergeevich}, {A. Chandran}, {J. Combes}, {S. D. Bartlett}, and {H. M.
  Wiseman}}}]{sergeevich11}
\bibinfo{author}{\bibnamefont{{A. Sergeevich}}},
  \bibinfo{author}{\bibnamefont{{A. Chandran}}},
  \bibinfo{author}{\bibnamefont{{J. Combes}}},
  \bibinfo{author}{\bibnamefont{{S. D. Bartlett}}}, \bibnamefont{and}
  \bibinfo{author}{\bibnamefont{{H. M. Wiseman}}}, \bibinfo{journal}{Phys. Rev.
  A} \textbf{\bibinfo{volume}{84}}, \bibinfo{pages}{052315}
  (\bibinfo{year}{2011}).

\bibitem[{\citenamefont{{F. Husz\'ar} and {N. M. T. Houlsby}}(2012)}]{huszar12}
\bibinfo{author}{\bibnamefont{{F. Husz\'ar}}} \bibnamefont{and}
  \bibinfo{author}{\bibnamefont{{N. M. T. Houlsby}}}, \bibinfo{journal}{Phys.
  Rev. A} \textbf{\bibinfo{volume}{85}}, \bibinfo{pages}{052120}
  (\bibinfo{year}{2012}).

\bibitem[{\citenamefont{{C. E. Granade} et~al.}(2012)\citenamefont{{C. E.
  Granade}, {C. Ferrie}, {N. Wiebe}, and {D. G. Cory}}}]{granade12}
\bibinfo{author}{\bibnamefont{{C. E. Granade}}},
  \bibinfo{author}{\bibnamefont{{C. Ferrie}}},
  \bibinfo{author}{\bibnamefont{{N. Wiebe}}}, \bibnamefont{and}
  \bibinfo{author}{\bibnamefont{{D. G. Cory}}}, \bibinfo{journal}{New J. Phys.}
  \textbf{\bibinfo{volume}{14}}, \bibinfo{pages}{103013}
  (\bibinfo{year}{2012}).

\bibitem[{\citenamefont{{C. Ferrie} et~al.}(2013)\citenamefont{{C. Ferrie}, {C.
  E. Granade}, and {D. G. Cory}}}]{ferrie13}
\bibinfo{author}{\bibnamefont{{C. Ferrie}}}, \bibinfo{author}{\bibnamefont{{C.
  E. Granade}}}, \bibnamefont{and} \bibinfo{author}{\bibnamefont{{D. G.
  Cory}}}, \bibinfo{journal}{Quant. Inf. Proc.} \textbf{\bibinfo{volume}{12}},
  \bibinfo{pages}{611} (\bibinfo{year}{2013}).

\bibitem[{\citenamefont{{N. Wiebe} et~al.}(2014{\natexlab{a}})\citenamefont{{N.
  Wiebe}, {C. E. Granade}, {C. Ferrie}, and {D. G. Cory}}}]{wiebe14b}
\bibinfo{author}{\bibnamefont{{N. Wiebe}}}, \bibinfo{author}{\bibnamefont{{C.
  E. Granade}}}, \bibinfo{author}{\bibnamefont{{C. Ferrie}}}, \bibnamefont{and}
  \bibinfo{author}{\bibnamefont{{D. G. Cory}}}, \bibinfo{journal}{Phys. Rev. A}
  \textbf{\bibinfo{volume}{89}}, \bibinfo{pages}{042314}
  (\bibinfo{year}{2014}{\natexlab{a}}).

\bibitem[{\citenamefont{{N. Wiebe} et~al.}(2014{\natexlab{b}})\citenamefont{{N.
  Wiebe}, {C. E. Granade}, {C. Ferrie}, and {D. G. Cory}}}]{wiebe14a}
\bibinfo{author}{\bibnamefont{{N. Wiebe}}}, \bibinfo{author}{\bibnamefont{{C.
  E. Granade}}}, \bibinfo{author}{\bibnamefont{{C. Ferrie}}}, \bibnamefont{and}
  \bibinfo{author}{\bibnamefont{{D. G. Cory}}}, \bibinfo{journal}{Phys. Rev.
  Lett.} \textbf{\bibinfo{volume}{112}}, \bibinfo{pages}{190501}
  (\bibinfo{year}{2014}{\natexlab{b}}).

\bibitem[{\citenamefont{{M. P. V. Stenberg} et~al.}(2014)\citenamefont{{M. P.
  V. Stenberg}, {Y. R. Sanders}, and {F. K. Wilhelm}}}]{stenberg14}
\bibinfo{author}{\bibnamefont{{M. P. V. Stenberg}}},
  \bibinfo{author}{\bibnamefont{{Y. R. Sanders}}}, \bibnamefont{and}
  \bibinfo{author}{\bibnamefont{{F. K. Wilhelm}}}, \bibinfo{journal}{Phys. Rev.
  Lett.} \textbf{\bibinfo{volume}{113}}, \bibinfo{pages}{210404}
  (\bibinfo{year}{2014}).

\bibitem[{M. P. V. Stenberg, K. Pack, and F. K. Wilhelm,
  arXiv:1508.04412()}]{stenberg15}
M. P. V. Stenberg, K. Pack, and F. K. Wilhelm, arXiv:1508.04412.

\bibitem[{\citenamefont{{B. L. Higgins} et~al.}(2007)\citenamefont{{B. L.
  Higgins}, {D. W. Berry}, {S. D. Bartlett}, {H. M. Wiseman}, and {G. J.
  Pryde}}}]{higgins07}
\bibinfo{author}{\bibnamefont{{B. L. Higgins}}},
  \bibinfo{author}{\bibnamefont{{D. W. Berry}}},
  \bibinfo{author}{\bibnamefont{{S. D. Bartlett}}},
  \bibinfo{author}{\bibnamefont{{H. M. Wiseman}}}, \bibnamefont{and}
  \bibinfo{author}{\bibnamefont{{G. J. Pryde}}}, \bibinfo{journal}{Nature
  (London)} \textbf{\bibinfo{volume}{450}}, \bibinfo{pages}{393}
  (\bibinfo{year}{2007}).

\bibitem[{\citenamefont{{G. Y. Xiang} et~al.}(2011)\citenamefont{{G. Y. Xiang},
  {B. L. Higgins}, {D. W. Berry}, {H. M. Wiseman}, and {G. J.
  Pryde}}}]{xiang11}
\bibinfo{author}{\bibnamefont{{G. Y. Xiang}}},
  \bibinfo{author}{\bibnamefont{{B. L. Higgins}}},
  \bibinfo{author}{\bibnamefont{{D. W. Berry}}},
  \bibinfo{author}{\bibnamefont{{H. M. Wiseman}}}, \bibnamefont{and}
  \bibinfo{author}{\bibnamefont{{G. J. Pryde}}}, \bibinfo{journal}{Nat.
  Photon.} \textbf{\bibinfo{volume}{5}}, \bibinfo{pages}{43}
  (\bibinfo{year}{2011}).

\bibitem[{\citenamefont{{H. Yonezawa {\it et al.}}}(2012)}]{yonezawa12}
\bibinfo{author}{\bibnamefont{{H. Yonezawa {\it et al.}}}},
  \bibinfo{journal}{Science} \textbf{\bibinfo{volume}{337}},
  \bibinfo{pages}{1514} (\bibinfo{year}{2012}).

\bibitem[{\citenamefont{{K. S. Kravtsov} et~al.}(2013)\citenamefont{{K. S.
  Kravtsov}, {S. S. Straupe}, {I. V. Radchenko}, {N. M. T. Houlsby}, {F.
  Husz\'ar}, and {S. P. Kulik}}}]{kravtsov13}
\bibinfo{author}{\bibnamefont{{K. S. Kravtsov}}},
  \bibinfo{author}{\bibnamefont{{S. S. Straupe}}},
  \bibinfo{author}{\bibnamefont{{I. V. Radchenko}}},
  \bibinfo{author}{\bibnamefont{{N. M. T. Houlsby}}},
  \bibinfo{author}{\bibnamefont{{F. Husz\'ar}}}, \bibnamefont{and}
  \bibinfo{author}{\bibnamefont{{S. P. Kulik}}}, \bibinfo{journal}{Phys. Rev.
  A} \textbf{\bibinfo{volume}{87}}, \bibinfo{pages}{062122}
  (\bibinfo{year}{2013}).

\bibitem[{\citenamefont{{A. Blais} et~al.}(2004)\citenamefont{{A. Blais}, {R.
  Huang}, {A. Wallraff}, {S. M. Girvin}, and {R. J. Schoelkopf}}}]{blais04}
\bibinfo{author}{\bibnamefont{{A. Blais}}}, \bibinfo{author}{\bibnamefont{{R.
  Huang}}}, \bibinfo{author}{\bibnamefont{{A. Wallraff}}},
  \bibinfo{author}{\bibnamefont{{S. M. Girvin}}}, \bibnamefont{and}
  \bibinfo{author}{\bibnamefont{{R. J. Schoelkopf}}}, \bibinfo{journal}{Phys.
  Rev. A} \textbf{\bibinfo{volume}{69}}, \bibinfo{pages}{062320}
  (\bibinfo{year}{2004}).

\bibitem[{\citenamefont{{M. Mariantoni {\it et al.}}}(2011)}]{mariantoni11}
\bibinfo{author}{\bibnamefont{{M. Mariantoni {\it et al.}}}},
  \bibinfo{journal}{Science} \textbf{\bibinfo{volume}{334}},
  \bibinfo{pages}{61} (\bibinfo{year}{2011}).

\bibitem[{For a more detailed description of swap spectroscopy, see also
  [16].()}]{swap}
For a more detailed description of swap spectroscopy, see also [16].

\bibitem[{\citenamefont{{H. Yuan} and {C. F. Fung}}(2015)}]{yuan15}
\bibinfo{author}{\bibnamefont{{H. Yuan}}} \bibnamefont{and}
  \bibinfo{author}{\bibnamefont{{C. F. Fung}}}, \bibinfo{journal}{Phys. Rev.
  Lett.} \textbf{\bibinfo{volume}{115}}, \bibinfo{pages}{110401}
  (\bibinfo{year}{2015}).

\bibitem[{\citenamefont{{M. West}}(1993)}]{west93}
\bibinfo{author}{\bibnamefont{{M. West}}}, \bibinfo{journal}{J. Roy. Stat. Soc.
  B Met.} \textbf{\bibinfo{volume}{55}}, \bibinfo{pages}{409}
  (\bibinfo{year}{1993}).

\bibitem[{\citenamefont{{N. J. Gordon} et~al.}(1993)\citenamefont{{N. J.
  Gordon}, {D. J. Salmond}, and {A. F. M. Smith}}}]{gordon93}
\bibinfo{author}{\bibnamefont{{N. J. Gordon}}},
  \bibinfo{author}{\bibnamefont{{D. J. Salmond}}}, \bibnamefont{and}
  \bibinfo{author}{\bibnamefont{{A. F. M. Smith}}}, \bibinfo{journal}{Radar and
  Signal Processing IEE Proc.-F} \textbf{\bibinfo{volume}{140}},
  \bibinfo{pages}{107} (\bibinfo{year}{1993}).

\bibitem[{J. Liu and M. West in, {\it Sequential Monte Carlo Methods in
  Practice}, edited by A. Doucet, N. Freitas, and N. Gordon (Springer, New
  York, 2001).()}]{liu01}
J. Liu and M. West in, {\it Sequential Monte Carlo Methods in Practice}, edited
  by A. Doucet, N. Freitas, and N. Gordon (Springer, New York, 2001).

\bibitem[{\citenamefont{{Y. Shalibo} et~al.}(2010)\citenamefont{{Y. Shalibo},
  {Y. Rofe}, {D. Shwa}, {F. Zeides}, {M. Neeley}, {J. M. Martinis}, and {N.
  Katz}}}]{shalibo10}
\bibinfo{author}{\bibnamefont{{Y. Shalibo}}}, \bibinfo{author}{\bibnamefont{{Y.
  Rofe}}}, \bibinfo{author}{\bibnamefont{{D. Shwa}}},
  \bibinfo{author}{\bibnamefont{{F. Zeides}}},
  \bibinfo{author}{\bibnamefont{{M. Neeley}}},
  \bibinfo{author}{\bibnamefont{{J. M. Martinis}}}, \bibnamefont{and}
  \bibinfo{author}{\bibnamefont{{N. Katz}}}, \bibinfo{journal}{Phys. Rev.
  Lett.} \textbf{\bibinfo{volume}{105}}, \bibinfo{pages}{177001}
  (\bibinfo{year}{2010}).

\bibitem[{\citenamefont{{T. Palom\"aki {\it et al.}}}(2010)}]{palomaki10}
\bibinfo{author}{\bibnamefont{{T. Palom\"aki {\it et al.}}}},
  \bibinfo{journal}{Phys. Rev. B} \textbf{\bibinfo{volume}{81}},
  \bibinfo{pages}{144503} (\bibinfo{year}{2010}).

\bibitem[{\citenamefont{{J. Lisenfeld} et~al.}(2010)\citenamefont{{J.
  Lisenfeld}, {C. M\"uller}, {J. H. Cole}, {P. Bushev}, {A. Lukashenko}, {A.
  Shnirman}, and {A. V. Ustinov}}}]{lisenfeld10}
\bibinfo{author}{\bibnamefont{{J. Lisenfeld}}},
  \bibinfo{author}{\bibnamefont{{C. M\"uller}}},
  \bibinfo{author}{\bibnamefont{{J. H. Cole}}},
  \bibinfo{author}{\bibnamefont{{P. Bushev}}},
  \bibinfo{author}{\bibnamefont{{A. Lukashenko}}},
  \bibinfo{author}{\bibnamefont{{A. Shnirman}}}, \bibnamefont{and}
  \bibinfo{author}{\bibnamefont{{A. V. Ustinov}}}, \bibinfo{journal}{Phys. Rev.
  Lett.} \textbf{\bibinfo{volume}{105}}, \bibinfo{pages}{230504}
  (\bibinfo{year}{2010}).

\bibitem[{J. Kennedy and R. Eberhart, {\it Proc. IEEE Int. Conf. Neural Netw.},
  Perth, 1995 (IEEE, Piscataway, New Jersey, 1995), Vol. 4, p.
  1942()}]{kennedy95}
J. Kennedy and R. Eberhart, {\it Proc. IEEE Int. Conf. Neural Netw.}, Perth,
  1995 (IEEE, Piscataway, New Jersey, 1995), Vol. 4, p. 1942.

\bibitem[{A. P. Engelbrecht, {\it Fundamentals of Computational Swarm
  Intelligence} (John Wiley \& Sons, England, 2005)()}]{PSObook}
A. P. Engelbrecht, {\it Fundamentals of Computational Swarm Intelligence} (John
  Wiley \& Sons, England, 2005).

\bibitem[{\citenamefont{{A. Hentschel} and {B. C.
  Sanders}}(2010)}]{hentschel10}
\bibinfo{author}{\bibnamefont{{A. Hentschel}}} \bibnamefont{and}
  \bibinfo{author}{\bibnamefont{{B. C. Sanders}}}, \bibinfo{journal}{Phys. Rev.
  Lett.} \textbf{\bibinfo{volume}{104}}, \bibinfo{pages}{063603}
  (\bibinfo{year}{2010}).

\bibitem[{\citenamefont{{A. Hentschel} and {B. C.
  Sanders}}(2011)}]{hentschel11}
\bibinfo{author}{\bibnamefont{{A. Hentschel}}} \bibnamefont{and}
  \bibinfo{author}{\bibnamefont{{B. C. Sanders}}}, \bibinfo{journal}{Phys. Rev.
  Lett.} \textbf{\bibinfo{volume}{107}}, \bibinfo{pages}{233601}
  (\bibinfo{year}{2011}).

\bibitem[{\citenamefont{{A. Hayes} and {D. W. Berry}}(2014)}]{hayes14}
\bibinfo{author}{\bibnamefont{{A. Hayes}}} \bibnamefont{and}
  \bibinfo{author}{\bibnamefont{{D. W. Berry}}}, \bibinfo{journal}{Phys. Rev.
  A} \textbf{\bibinfo{volume}{89}}, \bibinfo{pages}{013838}
  (\bibinfo{year}{2014}).

\bibitem[{\citenamefont{{E. Magesan} et~al.}(2012)\citenamefont{{E. Magesan},
  {J. M. Gambetta}, {A. D. C\'orcoles}, and {J. M. Chow}}}]{magesan15}
\bibinfo{author}{\bibnamefont{{E. Magesan}}}, \bibinfo{author}{\bibnamefont{{J.
  M. Gambetta}}}, \bibinfo{author}{\bibnamefont{{A. D. C\'orcoles}}},
  \bibnamefont{and} \bibinfo{author}{\bibnamefont{{J. M. Chow}}},
  \bibinfo{journal}{Phys. Rev. Lett.} \textbf{\bibinfo{volume}{109}},
  \bibinfo{pages}{080505} (\bibinfo{year}{2012}).

\bibitem[{\citenamefont{{F. Zhao} et~al.}(2013)\citenamefont{{F. Zhao}, {J.
  Tang}, {Ji. Wang}, {Ju. Wang}, {K. D. Osborn}, {A. Mizel}, {F. C. Wellstood},
  and {B. S. Palmer}}}]{zhao13}
\bibinfo{author}{\bibnamefont{{F. Zhao}}}, \bibinfo{author}{\bibnamefont{{J.
  Tang}}}, \bibinfo{author}{\bibnamefont{{Ji. Wang}}},
  \bibinfo{author}{\bibnamefont{{Ju. Wang}}}, \bibinfo{author}{\bibnamefont{{K.
  D. Osborn}}}, \bibinfo{author}{\bibnamefont{{A. Mizel}}},
  \bibinfo{author}{\bibnamefont{{F. C. Wellstood}}}, \bibnamefont{and}
  \bibinfo{author}{\bibnamefont{{B. S. Palmer}}}, \bibinfo{journal}{Appl. Math.
  Inf. Sci.} \textbf{\bibinfo{volume}{7}}, \bibinfo{pages}{499}
  (\bibinfo{year}{2013}).

\bibitem[{Y. Shi and R. Eberhart, in {\it Proceedings of the IEEE International
  Conference on Evolutionary Computation, Anchorage, Alaska, 1998} (IEEE, New
  York, 1998), p. 69()}]{shi98}
Y. Shi and R. Eberhart, in {\it Proceedings of the IEEE International
  Conference on Evolutionary Computation, Anchorage, Alaska, 1998} (IEEE, New
  York, 1998), p. 69.

\bibitem[{\citenamefont{{E. Zahedinejad} et~al.}(2014)\citenamefont{{E.
  Zahedinejad}, {S. Schirmer}, and {B. C. Sanders}}}]{zahedinejad14}
\bibinfo{author}{\bibnamefont{{E. Zahedinejad}}},
  \bibinfo{author}{\bibnamefont{{S. Schirmer}}}, \bibnamefont{and}
  \bibinfo{author}{\bibnamefont{{B. C. Sanders}}}, \bibinfo{journal}{Phys. Rev.
  A} \textbf{\bibinfo{volume}{90}}, \bibinfo{pages}{032310}
  (\bibinfo{year}{2014}).

\bibitem[{\citenamefont{{J. B. Chang {\it et al.}}}(2013)}]{chang13}
\bibinfo{author}{\bibnamefont{{J. B. Chang {\it et al.}}}},
  \bibinfo{journal}{Appl. Phys. Lett.} \textbf{\bibinfo{volume}{103}},
  \bibinfo{pages}{012602} (\bibinfo{year}{2013}).

\end{thebibliography}
\end{document}